# The passive operating mode of the linear optical gesture sensor

Krzysztof CZUSZYNSKI, Jacek RUMINSKI, Jerzy WTOREK
*Faculty of Electronics, Telecommunications and Informatics,
Gdansk University of Technology, 80-233,Gdansk, Poland*
krzysztof.czuszynski@pg.edu.pl

*Abstract*—The study evaluates the influence of natural light conditions on the effectiveness of the linear optical gesture sensor, working in the presence of ambient light only (passive mode). The orientations of the device in reference to the light source were modified in order to verify the sensitivity of the sensor. A criterion for the differentiation between two states: "possible gesture" and "no gesture" was proposed. Additionally, different light conditions and possible features were investigated, relevant for the decision of switching between the passive and active modes of the device. The criterion was evaluated based on the specificity and sensitivity analysis of the binary ambient light condition classifier. The elaborated classifier predicts ambient light conditions with the accuracy of 85.15%. Understanding the light conditions, the hand pose can be detected . The achieved accuracy of the hand poses classifier trained on the data obtained in the passive mode in favorable light conditions was 98.76%. It was also shown that the passive operating mode of the linear gesture sensor reduces the total energy consumption by 93.34%, resulting in 0.132 mA. It was concluded that optical linear sensor could be efficiently used in various lighting conditions.

## I. Introduction

The emergence of wearable smart devices has been stimulating research on the human system interaction methods across the decades [1–4]. The contactless navigation is a feature which especially allows devices to be utilized in a wide range of applications (e.g. healthcare, industry.) Non-contact interfaces based on video analysis are already popular also within mobile devices [5,6]. Even though they can handle a variety of gestures, they are computationally [5] and energetically expensive [7]. Thus what is of interest are sensors relying on a less robust computation that could handle a rich set of gestures with a high recognition accuracy along with a lower power consumption. Many of them are active sensors utilizing the excitation of a given type, such as optical [8–11] or radio wave sensors [12,13]. Yet passive solutions of gesture sensors based on many kinds of transducers have also been widely presented. The retransmission of captured WiFi signals and the echo signal reflected from a hand were utilized for gesture recognition by [14]. Their system utilizing the passive radar technology and relying on the Doppler effect is reported to detect five dynamic discrete gestures. The system was subsequently expanded to use LTE signals [15]. The wearable, glove-based system for tracking hand gestures with passive RFID sensor tags was proposed by [16]. The recognition of circular and semicircular gestures in three dimensions using the non contact sensor of passive RFID tags was investigated by [17]. The power saving sensor, able to recognize 8 gestures and utilizing wireless signals (e.g. TV, RFID) was presented in [18]. Capacitance changes in a three-electrodes set were utilized to track the position of a finger in two axes [19].

Other group of gesture sensors are optical sensors. 4x4 and 2x2 PIR sensor arrays were proposed for the detection of swipe gestures by [20,21]. The observation of ambient light modulations produced by a human hand was also utilized in a number of sensors. The user computing activity (keystrokes) was monitored with the use of ambient light sensors from a smart watch [22]. The optical passive sensor comprised of photodiodes arranged in a 3x3 array, utilizing only ambient light, was designed in [23]. As stated, as many as 10 dynamic discrete gestures were detected with a high accuracy. However, the sensor worked properly in neither very bright lights nor the dark.

A sensor operating in the passive mode does not use any excitation for measurement purposes (e.g. own light, radio waves). Therefore, it can save more power in comparison to a situation when additional, active electronic devices are used as a source of excitation. The passive mode of a gesture sensor operates in the existing environmental conditions, which can highly influence the measurement and the ability of an accurate recognition of gestures.

Optical sensors are the ones the performance of which may especially depend on ambient light level. Optical gesture sensors that can operate in either passive or active mode depending on ambient light conditions are not common in the literature. Therefore, the design of a sensor which could adapt the operating mode to the existing environmental conditions and preserve its gesture recognition capabilities would be of interest.

The goal of this research is to measure the behavior of the optical linear gesture sensor operating in the passive mode in different environmental conditions. Particular objectives for the evaluation of the passive operating mode are: to investigate the recognition accuracy of static poses; to propose the gesture / no gesture decision criterion; to evaluate the ambient light brightness range the optical linear sensor can reliably work in. The practical motivation for this work is the reduction of power consumption by using the passive mode of the sensor as often as possible. Additionally, the work also focuses on the investigation of the conditions when the gesture sensor could automatically choose the most suitable operating mode: passive or active.

The paper is organized as follows; Section I consists of the introduction, state of the art, the motivations and objectives of the work. Section II presents the description of the experiments and methods utilized to measure the properties of the optical sensor operating in the passive mode. Section III describes the results obtained from the experiments. The discussion of the results is presented in Section IV. The paper is concluded in Section V.

## II. MATERIALS AND METHODS

### A. The optical linear gesture sensor

The research was conducted on the prototype of the linear optical gesture sensor based on 8 IR photodiodes (Fig. 1). The applied elements (TSL260RD) are distanced from one another by 1 cm. The device is also equipped with 4 IR LEDs as a source of illumination for the active mode but given the scope of this paper, their utilization was not of interest. The light collimator part (black part in Fig. 2) of the sensor limits the field of view of photodiodes (PDs) and LEDs to 60° and 120°, respectively. It increases the spatial resolution of individual optical elements of the sensor [24]. The device is managed by the pic24FV16KA302 microprocessor (mounted on the bottom side of the PCB) and supplied with a 5V battery. The linear sensor is intended for the detection of hand gestures performed nearby the device, typically up to 5 cm (or up to 10 cm for wide reflecting objects) from the sensor plane [25]. Such a distance reduces potential interferences from other nearby objects, for example if the sensor is embedded in smart glasses (Fig. 2a) [26]. The sensor can be utilized for discrete commands (e.g. "next", "enter", "back") or continuous, mouse cursor like navigation (Fig. 2b).

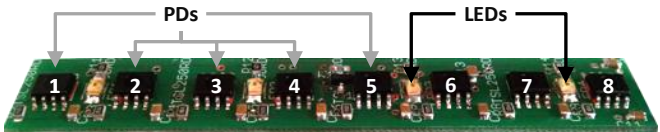

Figure 1. The linear optical gesture sensor utilized in the study with no overlay (light collimator) presented.

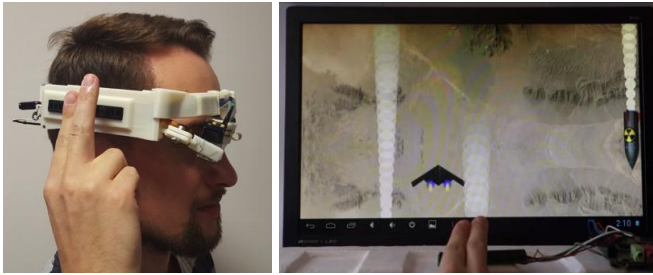

Figure 2. a) Person wearing the smart glasses equipped with the optical linear gesture sensor and performing the "2 fingers joined" (2FJ) hand pose. b) An example of the utilization of the sensor for a video game navigation.

### B. Principle of operation – operating modes

In the active mode, the LEDs of the sensor pulsate synchronously with the frequency of 40Hz [27]. In this mode 4 IR LEDs are the source of the light which reflects from a hand performing a gesture nearby the sensor. The intensity of the reflected light is measured by the aligned PDs and sampled by the microprocessor into a data frame (DF). The data frame contains 8 values measured by 8 photodiodes. This light intensity pattern is normalized by reducing all values in the DF by the *min*(DF) factor (Fig. 3a, 3b). Various finger arrangements (poses) produce reflection patterns which can be differentiated using a classifier, e.g. artificial neural network (ANN) used in [28].

In the passive operating mode solely ambient light is used, without any additional source of light (like IR LEDs). In this mode the light would be blocked by a hand performing a gesture producing a shadow pattern. After the inversion of the sampled shadow pattern its shape is similar to the light intensity pattern obtained in the active mode (Fig. 3c). The shadow patterns are normalized by inversion (multiplying by -1) and addition of the *min*(DF) obtained from the DF after the inversion (Fig. 3d). The source of the ambient light could be natural (sunlight) or artificial (e.g. room lights). In this study we analyze only the natural light.

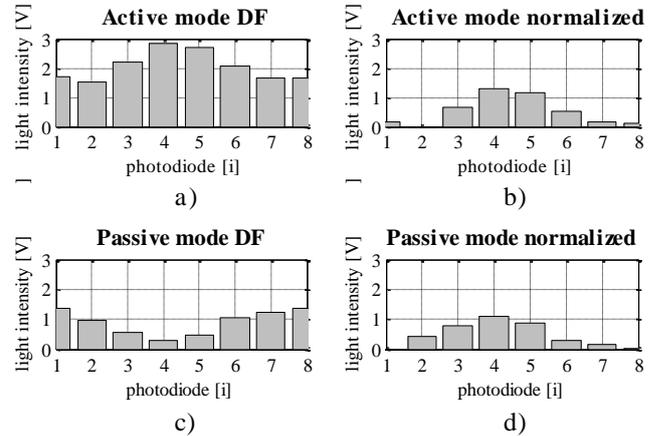

Figure 3. The pattern sampling performed during the cloudy day with two fingers joined pose located in front of the sensor. a) The DF sampled in the active mode. b) The normalized active mode DF. c) The DF sampled in the passive mode. d) The normalized passive mode DF.

### C. Objectives for experiments

The ambient sunlight level, its directivity and angle of incidence can impact the usability and reliability of the passive operating mode of the optical linear gesture sensor. Therefore, in this study we observe a few parameters related to the gesture detection capabilities of the sensor.

Not all recorded data frames should be used in further processing for the hand pose classification. For example, if all 8 values in DF are almost exactly the same, then it is highly probable that no gesture is being performed. Therefore, preprocessing of the data frames is performed to calculate the standard deviation (*sd*), *max*, *min* and their difference (*diff*) for each data fame. For example, in the active mode, when the value of the standard deviation of the 8 values in DF exceeds a given, experimentally established threshold value ($T_{sd}$), the analyses of the stored pattern are triggered. For the active mode, the $T_{sd}$ value was previously established as 0.1V [27]. However, the relevant threshold value for the passive mode may be different.

Other interesting parameters that should be evaluated from the data frames are: the longitudinal component of the localization of a hand in relation to the sensor and the code of the recognized pose of a hand. The localization of the hand can be estimated by calculating the center of gravity (*COG*) of the light pattern represented by the DF values [27]. The second parameter, the code representing the hand pose, is the output from the selected hand pose classifier. The classifier was designed to recognize 3 hand finger arrangements differing in the width of the plane produced by the fingers. These arrangements are "1 finger separated", "2 fingers joined" and "4 fingers joined" and the corresponding codes are 1FS, 2FJ and 4FJ, respectively.

Another important objective for the experiments is to investigate the influence of the environmental conditions on the output parameters of the gesture sensor working in the passive mode. In order to evaluate the acceptable ambient light conditions for the utilization of the passive mode of the gesture sensor, the following procedure is considered. The sensor is to be mounted on the table and rotated so as to measure the referential light characteristics (*max* parameter of the DF) of the room. Then, the hand pose mimicking obstacle would be hitched in front of the sensor, the rotary measurements would be repeated and the results would be normalized by the referential waveform. Additionally, any ambient light change would be recorded by a separate light meter. If the resulting normalized function correlated with the light level from the external light meter, it would mean that the *max* parameter changes with ambient light with no regard to the position of the sensor in reference to the sun direction. Thus it is a 2FJ pose that is to be selected as the one producing a pattern of moderate width (in relation to the width of the sensor) to be utilized in the measurements. Its shadow would most likely not cover the whole sensor and the obtained value of the *max* parameter would be close to the maximum value which would be recorded without the presence of an obstacle. In the positions (angles) where the classifier recognized the pose properly, ambient light conditions (represented by the *max* parameter) would be considered as acceptable. Wrong recognition is to be interpreted as unacceptable conditions.

*D. Datasets and ANN classifiers*

At the beginning, experiments with the participation of users were performed. Two types of datasets were recorded: datasets obtained for the active mode of the sensor and datasets for the passive mode of the sensor.

The active mode dataset consisted of 6600 samples from 11 volunteers (mean age 31 years; 7 males and 4 females), each producing 200 data frames of 3 gestures. The data was gathered with no presence of ambient light.

The passive mode dataset was gathered in four ambient light conditions, possibly linearly spaced due to the brightness level. The average brightness during the data collection sessions was 230, 700, 1460 and 2200 lux. The set of 6000 samples was collected by three volunteers (mean age 29 years; all subjects were males), each producing 500 data frames of 3 gestures in one session. The measurements were performed with the face of the sensor perpendicular to the light source direction (favorable conditions).

Each dataset was divided into training, validation and testing subsets. First, the Artificial Neural Network (ANN) classifier was learned in the Matlab software on the training subset of the active mode dataset. The obtained model will be later referred to as *aANN* [28]. Next, the ANN classifier was learned on the training subset of the passive mode dataset. The obtained model will be later referred to as *pANN*. Both models were compared using test subsets of the datasets.

*E. Laboratory setup*

The laboratory setup was built so as to measure the performance of the sensor in varied light conditions. The rotation of the sensor in relation to the direction of the sun changes the sunlight incident angle as well as the measurement conditions. Therefore, the measurements were performed utilizing the constructed stand with a rotary holder. According to the spherical coordinate system, the rotations of the sensor in φ and θ angles could be performed as illustrated in Fig. 4.

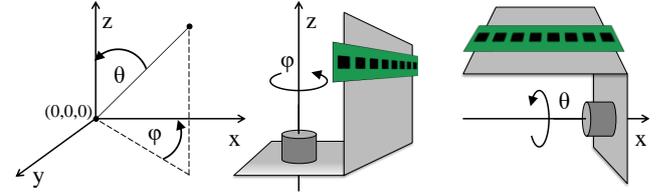

Figure 4. The scheme of the relation between the rotation axes and position of the linear sensor (elongated rectangle) in φ and θ rotation measurements.

The sensor hitched to the rotary holder in the position ready for performing the φ rotation experiments is presented in Fig. 5. A cardboard screen of the shape of a human head was attached to the back of the sensor. The purpose of this application is to replicate the scenario of the sensor built in the frame of smart glasses where the head covers the back of the sensor. In the experiments, the holder with the sensor was mounted on a rotary platform on a table, 1 m above the ground. The hemi-sphere probe of the light meter was placed on the same table, faced up. The measurements were performed for natural day light in different conditions.

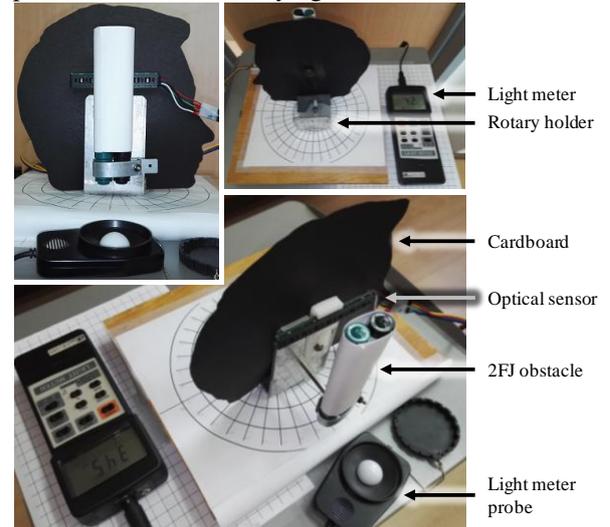

Figure 5. The description of the laboratory setup. A 2FJ pose mimicking obstacle is hitched in front of the sensor.

During the experiments, the holder was placed in the center of a room, in its starting position being faced towards the only window (southern side) and with the longer side of the device parallel to the ground. At each of the following positions (angles), the DF was sampled and data were send to the PC by the UART interface of the sensor. The ambient light level was monitored utilizing the Lutron LX-105 light meter and saved to the PC using the RS232 interface.

Additionally, the removable hitched obstacle mimicking the two fingers joined pose (2FJ)[24] can be attached to the rotary holder. It was utilized in order to observe the impact of varied ambient light conditions on the parameters of the DF of the stable sensor-hand finger unit. The width of artificial fingers (32 mm) was taken upon the study performed on the group of volunteers [27]. A 2FJ obstacle was mounted in front of the center of the sensor, 20 mm from the face of the PDs (around the half of the assumed operating distance [27]).

*F. Laboratory experiments*

*1) Angular characteristics measurements*

In φ angle variation experiments, the rotary holder with the sensor was rotated clockwise with the step of 10° performing a full circle (36 positions). In θ angle variation experiments, the holder was rotated up to the ceiling and then down to the back, with the step of 10°, performing a half circle (19 positions). Each of the angular experiments was performed in 4 scenarios. In the first two ones, the sensor was measuring the characteristics of the room – the "no obstacle" measurement – for weak (100 – 400 lux) and strong (600 – 2000 lux) ambient light conditions. The half of the saturation level of the PDs is 1.9V, which was measured to be 592 lux for the sunlight, hence the border level between the conditions. In the third scenario, ambient light was within the weak light range but a 2FJ obstacle was hitched in front of the sensor. In the fourth scenario, the obstacle was present as well but ambient light was in the stronger range.

*2) Distal characteristics measurements*

The ratio of components of the directed and scattered light can have an impact on the pattern of shadow produced by an obstacle, thus influencing the parameters calculated by the sensor. The phenomenon can be even more prominent given that the distance between an obstacle and the sensor changes. Therefore, the measurements with the artificial 2FJ obstacle initially located in front of the sensor at the distance of 1 cm and shifted up to 10 cm with a step of 1 cm were conducted for differentiated ambient light conditions. The measurements were performed in three scenarios, all with a 2FJ obstacle hitched. Apart from the weak and stronger light conditions described in the previous paragraph, there was also a very dark light scenario (below 100 lux).

*3) Power consumption*

In the passive operating mode, the only components of the total power consumption of the optical linear gesture sensor are the current drawn by 8 photodiodes and the current of the microcontroller. In this paper we focus on the power requirements of the applied PD chips (the transducer part of the sensor). According to the catalogue note of the applied PDs, their supply currents may vary. Therefore, the current consumption of the optical elements can be measured for different levels of ambient light.

## III. RESULTS

*A. Performance of the ANNs*

The ANN for the recognition of hand poses with joined fingers using data recorded by the active optical linear sensor was first developed and described in [28]. In this work the model was trained with the *active mode dataset* using the following settings: the training/validation/testing subsets ratio equal to 0.7/0.15/0.15; 1 hidden layer considered with up to 30 neurons; top 9 features selected by the matrix of correlation coefficients. The selected features of the pattern included: full width at 50% of *max*, full width at 85% of *max*, *COG, mean*, angle (slope of the pattern), *sd*, kurtosis, number of values in DF smaller than 2·*sd* and number of values in DF greater than the *mean*. The most efficient topology of the network consisted of 22 hidden layer neurons. The resulting classification accuracy of the *aANN* was 93.46% in comparison to 90.02% obtained earlier in [28]. The *aANN* was then tested on the *passive mode dataset,* and the resulting accuracy was 75.51%.

The set of features from the *passive mode dataset* was extended by the parameter *rawmax*, which describes the maximum of the obtained pattern before the normalization. Utilizing the same elimination method (matrix of correlation coefficients), the top 9 features were selected. The selected features were: full width at 15% of *max*, full width at 85% of *max*, *COG*, *mean*, angle (slope of the pattern), skewness, kurtosis, number of values in DF smaller than 2·*sd* and *rawmax*. The other settings were the same as applied in the learning of the *aANN*. The most efficient topology, with 25 neurons in the hidden layer, has the accuracy of 98.76%. The *pANN* was also evaluated on the *active mode dataset* resulting in scarcely 52.82% recognition rate. The cross-checking was performed because the active/passive profiles after the normalization were similar, therefore it was interesting to perform the cross tests (e.g., *aANN* with *passive mode test subset* and *pANN* with *active mode test subset*). The summary of the results is presented in Table I.

TABLE I. THE ACCURACY OF THE CLASSIFIERS TRAINED ON THE ACTIVE AND PASSIVE DATA EVALUATED ON SUBSETS OF DIFFERENT ORIGIN

|  |  |  | Test subset type | |
|---|---|---|---|---|
|  |  |  | Passive | Active |
| classifier | pANN | Train subset type | Passive | 98.76 % | 52.82 % |
|  | aANN |  | Active | 75.51 % | 93.46 % |

*B. No gesture threshold value*

The hardware of the gesture sensor, all of the decisions and analysis performed within the implemented logic of its operating system can be, in overall, named as the architecture of the Gesture Recognition System (GRS) controller [29]. In the research on the active operating mode of the linear gesture sensor described in [28], a dedicated architecture of the GRS was proposed. One of its tasks was to decide based on the *sd*(DF) value whether the data frame should be analyzed or "no obstacle" state is to be expected. Considering the possible utilization of the passive operating mode, the preceding decision can also be introduced to such GRS. As long as ambient light conditions are fine for the passive mode, the LEDs should be turned off.

In the previous research, with the active operating mode and no ambient light considered, the threshold of *sd*(DF) was set to 0.1V [27,28]. Nevertheless, this value may be inadequate in the presence of ambient light. For the four angular measurements with no obstacle in front of the sensor (rotations in φ, θ angles in the presence of weak and strong light), the value of *sd* of the DF in 110 (2·36+2·19) positions was measured. The ambient light value was in the range from 229 to 1108 lux. The obtained span of the *sd* values was from 0.0052 to 0.32V. For the $T_{sd}$ threshold of 0.1V, only 82.72% of the data frames were correctly classified as "no obstacle" (19 samples were above the limit) despite the calibration of the photodiode sensitivity functions. However, bringing the $T_{sd}$ just to 0.13V increases the classification accuracy to 94.54%. This value was applied in the analysis in the following experiments in the paper. The more precise adjustment of this value can be investigated in the future.

## C. Angular and distal characteristics

Each type of the measurement (angular or distal, Fig. 7-9) with an obstacle was conducted in two types of ambient light conditions (weak and strong light) and performed three times. Additionally, for the angular measurements there were also measured the characteristics with "no obstacle" hitched. The results of the exemplary measurements are presented in the subsections C and D. The selected measurements have the central brightness recorded by the light meter among the three measurements performed in a given configuration.

### 1) The rotation in φ angle

The first part of the experiment was performed during the cloudy day, with no obstacle in front of the sensor (reference). The average ambient light level during the measurement was 282±53 lux. Next, a pose imitating obstacle was hitched in front of the sensor and the rotations were repeated. The average measured brightness was 256±21 lux. The measurements with the obstacle were performed also at the other day, during sunny conditions. The average measured brightness was 1308±144 lux. The red zone on the polar plots of standard deviation of the DF (Fig. 7a, c, e) represents the angles for which the $sd$ was below the threshold, $T_{sd}$. The green zone (if present) helps to emphasize positions at which the sensor notices an obstacle ($sd(DF) > T_{sd}$).

The radius of the green zone is the average value of the $sd(DF)$ during the experiment (all positions). The color lines on the amplitude related plots (Fig. 7b,d,f) represent the $max$, $mean$ and $min$ value of the DF during the rotations. The $diff$ parameter is represented by a colored zone in order to emphasize it on the plot. The used PD chips of the sensor saturate at the level around 3.8V, hence the value of the $max$ parameter can be limited.

### 2) The rotation in θ angle

The first part of this experiment was performed during the cloudy day, with no obstacle in front of the sensor. The average ambient light level during the measurement was 229±2 lux. Next, a pose imitating obstacle was inserted in front of the sensor. The average measured brightness was 221±16 lux. At the other day, during sunny conditions, a pose imitating obstacle was hitched in front of the sensor. The average measured brightness was 837±63 lux. The plots of standard deviation of the DF are presented on Fig. 8a, c, e while the $diff$ parameter for different light conditions is presented in Fig. 8b, d, f.

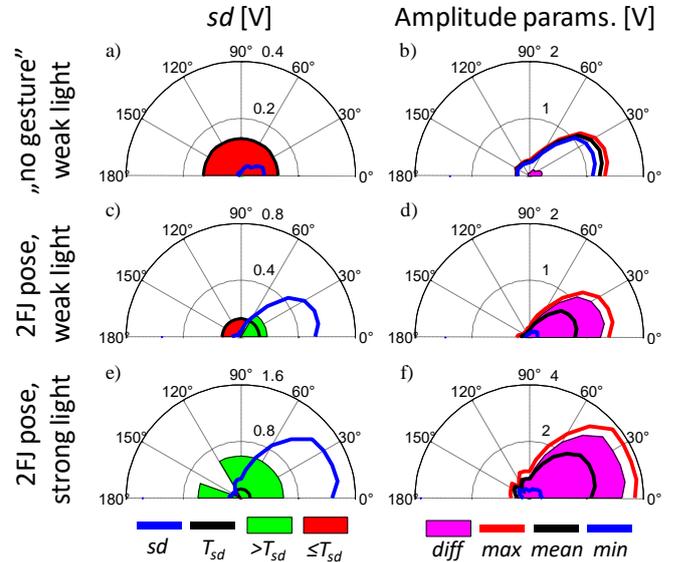

Figure 8. Results for the θ rotation. Side view, 0° is the window direction. The "no obstacle" angular characteristics of the sensor during the cloudy day (a, b), with a 2FJ obstacle (c, d), the sunny day with an obstacle (e, f).

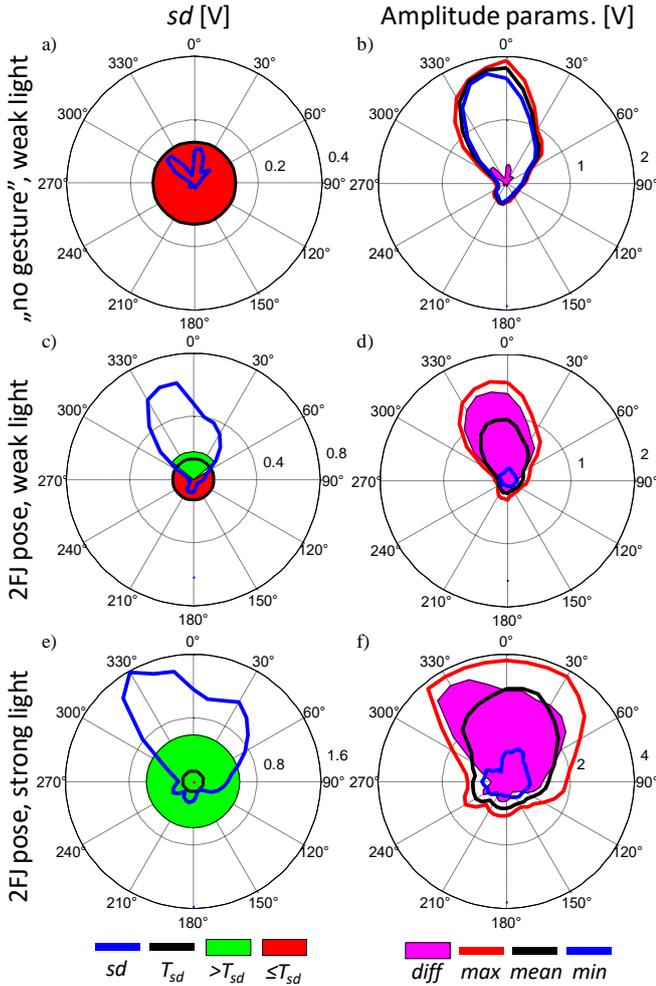

Figure 7. Results for the φ rotation. Top view, 0° is the window direction. The angular characteristics of the sensor during the cloudy day with no obstacle (a, b), with a 2FJ obstacle on the cloudy day (c, d), during the sunny day with a 2FJ obstacle (e, f). The blue plots at a, c, e denote $sd$, whereas b, d, f stand for amplitude related parameters.

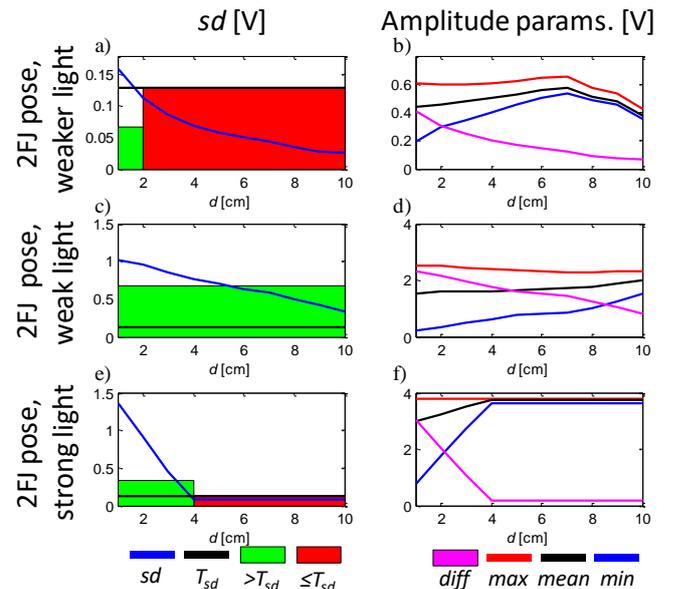

Figure 9. The distal measurements for a 2FJ obstacle. The characteristics taken during the sunny day with blinds on the window and sensor turned back (a, b), turned to the window with blinds (c, d), without blinds (e, f).

*3) Distance increase*

In all of the distal characteristics experiments, a 2FJ obstacle was hitched in front of the sensor. The first part of the experiment was performed during the sunny day, with blinds on the window and with the sensor turned back to the window. The average ambient light level in the room during the measurement was 401±60 lux. The second part of the experiment was performed also with the blinds but with the sensor faced to the window. The average ambient light level during the measurement was 426±35 lux. In the last measurement, the blinds were removed with the sensor still faced to the sunlight direction. The average measured brightness was 1838±34 lux, more than for any rotation based experiments. The results are presented in Fig. 9. Parameter $d$ is the distance between the obstacle and the face of the sensor, which varied in this experiment.

## D. Gesture recognition capabilities

For the measurements with an obstacle presented in the previous section, the results of the further analysis are presented. The gesture recognition related parameters were calculated for the positions where the $sd>T_{sd}$.

*1) The rotation in φ angle*

The *pANN* pose classifier was utilized in order to detect a static pose (2FJ) in differentiated ambient light conditions. The bold dashed line on the ANN polar plots (Fig. 10a, b) represents the expected class among the three ones which the classifier was trained on. A 2FJ pose in the weak light was recognized in 100% of the positions where the $sd$(DF) was greater than $T_{sd}$. In the strong light, the recognition rate was 92.67%.

The bold dashed line on the COG plots indicates the expected calculated position of a pose in relation to the sensor (Fig. 10c, d). In order to observe absolute calculated position errors in both directions, the 0 cm (referring to the center of the sensor) is not located in the centre of a plot.

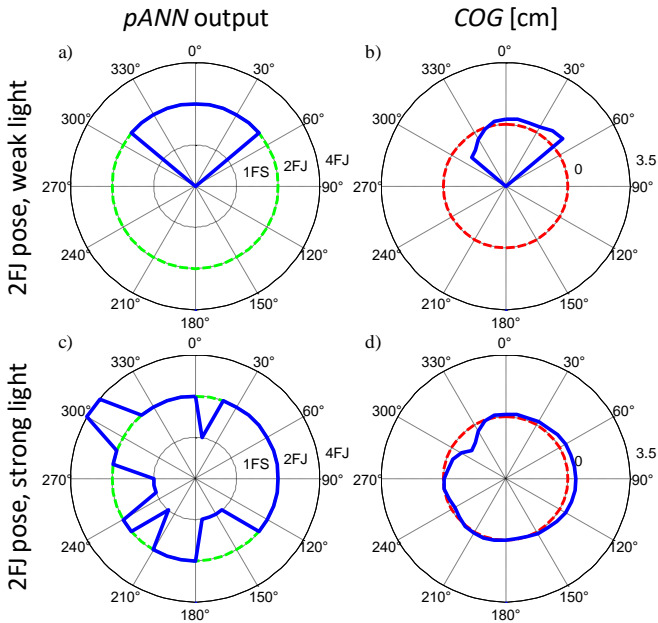

Figure 10. Results for the φ rotation. Top view, 0° is the window direction. The angular characteristics of the sensor during the cloudy day with an obstacle (a, b), and during the sunny day with a 2FJ obstacle (c, d). The plots a, c present the class recognition by ANN, where the dashed line is the expected result; plots b, d denote the calculated COG, with the dashed line being the expected value.

*2) The rotation in θ angle*

In the presented exemplary measurements of the θ rotation, the *pANN* classifier recognized a pose at all of the positions taken into analysis (Fig. 11a, c). The corresponding *COG* parameter values were calculated in the same range (Fig. 11b, d).

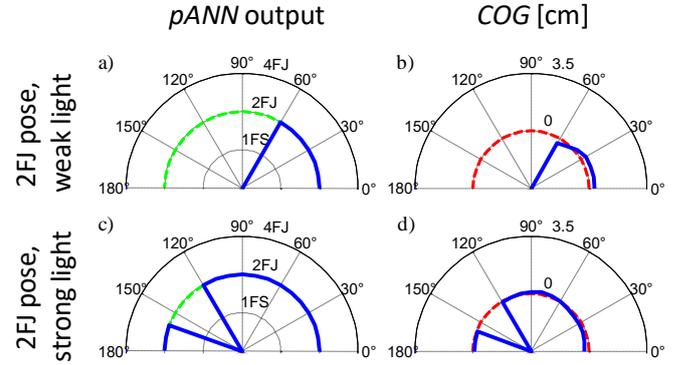

Figure 11. The θ rotation. Side view, 0° is the window direction. The angular characteristics of the sensor during the cloudy day with an obstacle (a, b), and during the sunny day with a 2FJ obstacle (c, d). The plots a, c denote the class recognition by ANN, whereas the dashed line is the expected result; plots b, d present the calculated COG, with the dashed line being the expected value.

*3) Distance increase*

In the exemplary distal characteristics the *pANN* classifier recognized a pose with 100% accuracy in weak light conditions. In very strong light the pose was recognized properly only at the closest distance of 1 cm (Fig. 12a, c). The corresponding COG parameter values in both cases were calculated (Fig. 12b, d).

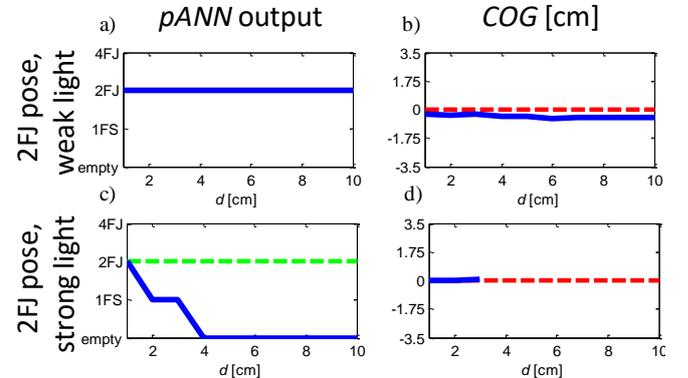

Figure 12. The distal measurements for a obstacle. The characteristics taken during the sunny day with blinds on the window and the sensor turned to the window (a, c) and without blinds (b, d).

## E. Summary of rotation measurements

As stated, each of the rotation based measurements was performed three times for each type of ambient light conditions. Therefore, the sets for analyses (for weak and strong light) from the φ angle rotations consisted of 108 samples. The analysis sets from the θ angle rotations consisted of 57 samples. These experiments are more meaningful for the purpose of this article than the distal characteristics since they include more information on the impact of varied light (angle, intensity) on the pose classification accuracy. Therefore no summary of the distal characteristic is presented. The results of the brightness measured during the experiments, visibility ranges of the sensor, accuracies of the *pANN* classifier at positions where $sd(\text{DF})>T_{sd}$ as well as the values of COG are presented in Table II.

TABLE II. SUMMARY OF THE MEASUREMENTS PERFORMED IN DIFFERENT CONFIGURATIONS (LIGHT CONDITIONS AND ROTATION ANGLE)

| Config. | brightness [lux] | range [°] | pANN acc. [%] | COG [cm] |
|---|---|---|---|---|
| φ rotation, weak light | 278.1±18.2 | ±56.7±2.9 | 100±0.0 | 0.08±0.56 |
| θ rotation, weak light | 215.3±9.5 | ±31.7±5.8 | 84.2±24.7 | 0.13±0.21 |
| φ rotation, strong light | 1263±168 | ±158.3±25.7 | 75.8±13 | 0.02±0.36 |
| θ rotation, strong light | 792.2±59.1 | ±63.3±17.6 | 100±0.0 | 0.07±0.14 |

*F. Operating mode estimation*

For one of the "no obstacle" characteristics in θ angle, the standard deviation of the ambient light measured by the light meter was very low (below 2 lux). Therefore, the obtained waveform was treated as the characteristic of the room at that time and for each of the 19 positions the max parameter of data frames was found. The resulting vector was normalized and as a result, the reference vector representing the given sunlight conditions for the room was obtained. The next three repetitions of the θ measurement, conducted briefly after the reference, were performed with a obstacle hitched into a holder. The *max* parameters obtained from the data frames were divided by the reference vector. The obtained values were then plotted against the ambient light brightness inside the room during the measurement, which was registered by the light meter (Fig. 13).

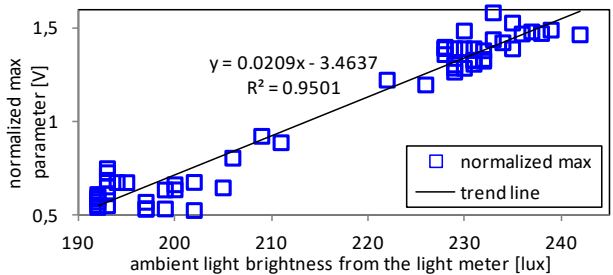

Fig. 13. *Max* parameter divided by the reference vector plotted against the ambient light brightness.

The obtained results suggest that the shadow produced by a obstacle does not affect the *max* vs. ambient brightness relationship as long as a obstacle is present in front of the sensor. Therefore, the *max* parameter was taken to represent ambient light conditions.

Each of the angular measurements with a obstacle presented in Section IIIC was additionally repeated twice in similar ambient light conditions. Therefore, the set of 6 φ and 6 θ measurements in total (each consisting of 3 strong and 3 weak light conditions) with a 2FJ obstacle hitched was taken into analysis on the active/passive mode switching criterion. At each position (angle) where the *pANN* classifier responded correctly, the corresponding parameter was assigned with the *bright* label. At the positions where the wrong call was noticed, the assignment with *dark* label was made. The *too bright* label was assigned to the errors which occurred at the saturation level of the PDs (around 3.8V).

The J48 classifier was applied utilizing the Weka software in order to evaluate the conditions at which the sensor is the most likely to operate in the passive mode (*bright*), active mode (*dark*) and whenever the brightness is too high (*too bright*). In order to keep the switching condition simple (fast computation), single parameter criterions were considered. The *max* parameter was automatically selected with the threshold value, $T_{max}$=0.387V, which divided the input dataset into *bright* and *dark* classes. No rule was produced for the class *too bright* as in the training set there were more correctly classified data frames with the *max* parameter in the saturation region than assigned representations of the class *too bright* overall. The obtained classifier recognizes the *bright* class samples with the accuracy of 100% while the *dark* class samples are recognized with 72.8%. The total accuracy of the classifier is 86.36%.

However, there is an asymmetry in the consequences of the detection accuracies of the classes *dark* (active mode recommended) and *bright* (passive mode recommended). The *dark* class error has to be minimized because the wrong decision of not turning on the illumination system (i.e., LEDs of the device) in *dark* light conditions can result in not detecting a gesture at all (the obtained pattern would be too flat, e.g. Fig. 9a, b). On the other hand, the samples of the *bright* class incorrectly classified as *dark* would lead to turning on the illumination system too early. At the cost of decreasing the utilization range of the passive operating mode, the poses can be still detected in the active mode in certain situations, though. However, the value of the baseline caused by the ambient light would have to be lower than the intensity of the LEDs light reflected from the hand.

By skipping the *too bright* class, the problem was redesigned as a binary classification problem. The true *dark* class samples classified as *dark* represent the True Positive samples in the binary classifier while the true *bright* class samples classified as *bright* are the True Negative samples. Therefore, we can observe the impact of adjusting the $T_{max}$ threshold ranging from 0 to 3.8V on the sensitivity and specificity of the binary classifier. It is presented on the ROC curve (Fig. 14a). The adjustment step was 0.01V.

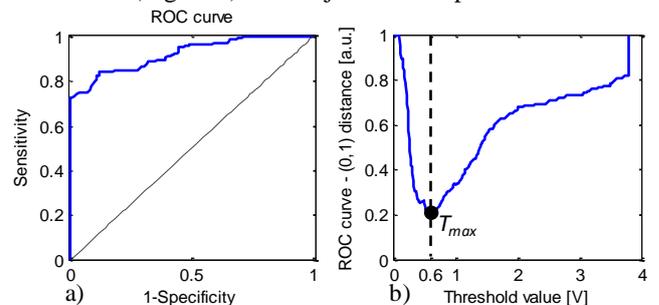

Figure 14. a) The ROC curve. b) The optimal ROC curve point plot.

The distance between each point of the ROC curve and the perfect binary classifier (0,1) was calculated (Fig. 14b). The closest point which indicates the optimum was achieved for the $T_{max}$=0.6V. This value corresponds to 142.7 lux measured by the light meter probe directed to the light source at the same angle as the photodiodes. The value was calculated from the linear function obtained from the linear regression analysis of the *max* vs. light brightness from the light meter.

After the application of the new, tuned threshold, the matrix of confusion can be obtained (Table III). Therefore, the sensitivity of the classifier is 84.11% while the specificity is 88%. Before the adjustment these values were equal to 69.54% and 100%, respectively. The overall accuracy of the binary light condition classifier is 85.15%.

TABLE III. THE CONFUSION MATRIX FOR OPERATING CONDITIONS AFTER THE CLASSIFICATION THRESHOLD ADJUSTMENT

| samples | Classified as *dark* | Classified as *bright* | |
|---|---|---|---|
| True *dark* | (TP) 127 | (FN) 24 | Sensitivity = 127/151 = 84.11% |
| True *bright* | (FP) 21 | (TN) 154 | Specificity = 154/175 = 88% |

*G. Power consumption*

The optical system component of the current consumption was measured in four situations: for no light, for two cases of bright dispersed light and with direct sunlight incident on the PDs. The current drawn from 8 PDs was averaged and the single photodiode current consumption for different light conditions is given in the column 3 of Table IV.

Based on the previous research, the microcontroller samples the analogue signals from the PDs at the rate of 40Hz [27]. The fill factor, $D$, indicates for how long during the sampling cycle (25ms) the optical elements (photodiodes or LEDs for the active mode) are supplied. Taking into account the measured settling time of the elements (375μs), the resulting $D$ is equal to 1.5%. It allows estimating the total current and power drawn by the optical system of the sensor. The results are presented in Table IV.

TABLE IV. THE CURRENT UTILIZATION OF THE PHOTODIODES OF THE SENSOR FOR DIFFERENT AMBIENT LIGHT CONDITIONS

| Light conditions | Brightness [lux] | Av. single PD current ($D$=100%) [mA] | Total (8) PD current ($D$=1.5%) [μA] | Total (8) PD power ($D$=1.5%) [μW] |
|---|---|---|---|---|
| dark | 0 | 0.959 | 115.05 | 575.25 |
| strong | 872 | 1.125 | 135.00 | 675.00 |
| stronger | 2000 | 1.235 | 148.20 | 741.00 |
| dir. sun. | 33500 | 1.041 | 124.95 | 624.75 |
| average | - | 1.090±0.12 | 130.80±14 | 654.00±71 |

IV. DISCUSSION

The evaluation of the *aANN* classifier using a testing subset from the *passive dataset* gave a low classification accuracy. Therefore a dedicated *pANN* classifier was trained and the resulting classification accuracy of the testing subset of the *passive dataset* has increased from 75.51% to 98.76%. Therefore, depending on ambient light conditions, not only the operating mode of the sensor should be chosen, but also the associated classification model (ANN classifier) has to be utilized to maximize the performance.

The analysis on a possible threshold value to differentiate between the gesture / no gesture states perceived by the optical sensor was performed. It was based on the four angular measurements in φ angle performed for different ambient light conditions and without any obstacle in front of the sensor face. Generally, in the darkness or very low lights, the $sd$(DF) value is low as almost no light reaches the sensor. In the passive mode, the impact of ambient light and non equal sensitivity of PDs may lead to larger and varied values of $sd$, even for the measurements not interrupted by the obstacle. A threshold $T_{sd}$ equal to 0.13V was experimentally chosen as almost 95% of the data frames from the four angular measurements (each representing the "no gesture" class) had a smaller value of the standard deviation. It means that in some rare circumstances the sensor may start to analyze and classify "empty" data frames (i.e., when no obstacles were present in front of the sensor face). Further research is required to improve the rejection rate of "empty" data frames.

The sensitivity of the sensor in different light conditions was measured by introducing a controlled change of the orientation of the sensor in reference to the light source. The measured light patterns for each orientation, after the application of the threshold value, $T_{sd}$, were marked as obtained in appropriate or inappropriate light conditions for the sensor. The light conditions were acclaimed as appropriate if the sensor operating in the passive mode had classified the obtained pattern caused by the hitched obstacle as a 2FJ pose. During the experiments, the sensor was centrally placed 2.6 m from a 3.1 m wide window, hence direct outside light was expected within the range of approximately ±30°. The results of the experiments show the average range at which (based on three measurements) the value of $sd$(DF) was greater than $T_{sd}$. For weak ambient light, the sensor recognized (but not necessarily correctly classified) any pose in the range of φ angles around ±57° (278±18 lux) and for the range of θ angles around ±32° (215±10 lux). For stronger light, these values rose to ±158° (1263±168 lux) and ±63° (792±59 lux), respectively. The degrees are given in reference to the light source (sun) direction. If the sensor were used within smart glasses, then the rotations in φ angle would be more natural and more frequent than in θ angle. In a simplified estimation, for a user who moves his/her head uniformly towards different directions in relation to the sun throughout the day, the obtained φ range can be averaged to ±107.5°. In reference to a full circle, it can give a rough estimation of saving the power (switching the sensor from the active to the passive mode) 60% of the time during a day.

In the estimation of a pose recognition accuracy, only outputs of the *pANN* at selected positions were taken to the statistics. The positions were selected whenever the threshold $T_{sd}$ was exceeded. The measurement for the given angle and kind of ambient light (dark / bright) was repeated three times and the results were averaged. For weak light conditions the accuracy was 84.2% for the θ rotations and 100% for the φ rotations. In the stronger light, the accuracies were at the level of 100% and 75.8% for the rotations in θ and φ angles. On average, these values differ from the very high recognition accuracy achieved on the testing subset from the *passive dataset*. The reason why it happens so may be that they were gathered for the differentiated orientations of the sensor and light source direction, while the testing subset was obtained in the favorable conditions (sunlight perpendicular to the sensor).

There are many works which describe the technology that can be utilized for passive gesture sensing but do not yet include research on the gesture detection accuracy [16,17,19]. Some pieces of research present varied gesture detection rate. The 2x2 PIR sensor recognizes 4 motion trajectories with the accuracy of 77% [21]. The AllSee sensor has the accuracy of 97% for the set of 8 gestures when utilizing the RFID signals [18]. The optical 3x3 gesture sensor described in [23] is reported to recognize 10 gestures with the accuracy of 98%. The classifier of this sensor was trained and evaluated for differentiated ambient light conditions, which were divided into two general categories, namely light and dark. However, the impact of incident ambient light angle changes on the gesture recognition accuracy was not investigated in detail in this paper. It is important to emphasize that the numbers presented for the linear optical sensor regard the detection of hand poses. The sensor still needs a higher level classifier which would allow to build a set of gestures based on the estimated hand localization and pose of a hand in subsequent sampling cycles [28].

An accurate detection of the position of hand fingers in reference to the sensor face for different lighting conditions could be very important for dynamic gesture recognition. Two parameters were investigated: longitudinal position of a hand determined by *COG* and distance from the sensor face. The x position was calculated for an obstacle, whereas the expected value was 0 cm (a pose above the center of the sensor). The average value of the *COG* parameter for the φ rotations in weak and strong lights were 0.08±0.56 cm and 0.02±0.36 cm, respectively. The average values of the calculated position of an obstacle in θ angle in weak and strong ambient light conditions were 0.13±0.21 cm and 0.07±0.14 cm, respectively. The significantly lower standard deviation of the position in the θ rotations can be attributed to the fact that the angle between the elongated part of the obstacle and the light source direction did not change in this experiment. The observable shifts to positive and negative values of the *COG* in the φ rotations are caused by the shift of the shadow of an obstacle. It demonstrates that the calculated position of fingers in the passive mode strongly depends on the mutual location of the sensor and the light source. However, the feature of continuous hand pose localization in relation to the sensor, as in [19], makes the linear gesture sensor an attractive solution among other basic sensors. It is a considerable advantage over some sensors that support only discrete gestures [9,20,23].

The pose recognition accuracy in the distal characteristics performed in weak light and in favorable conditions (sensor perpendicular to light source direction) showed that a pose can be recognized even from the distance of 10 cm. On the other hand, very strong light (above 1800 lux) causes that at the distance of 4 cm the shadow pattern becomes very weak and all of the PDs saturate (*diff* parameter goes to 0V). The recognition of the pose in such strong light occurred to be possible only at the distance of 1 cm. Therefore, the elaborated passive mode classifier for the linear sensor is not a reliable solution for very strong lights.

Initially, in order to find the optimal condition for switching between the passive and active operating modes of the sensor, the J48 classifier from the Weka software was applied. The ambient light conditions were represented by the *max*(DF) parameter and were classified as one of the three classes: *dark*, *bright* and *too bright*. The optimal threshold $T_{max}$ was equal to 0.387V and divided the light conditions with the accuracy of 86.36%. However, due to the asymmetry in the consequences of wrong classification of classes *dark* and *bright*, the sensitivity and specificity analysis was utilized. Additionally, the *too bright* class was rejected from the investigation as there were more correctly classified samples with the *max* parameter in the saturation region than the overall number of assigned samples of the class *too bright*. Therefore, the problem was solved for the binary classifier. The new threshold value of $T_{max}$ was automatically found as 0.6V and the sensitivity and specificity of the new classifier were 84.11% and 88%, respectively. In situations where the *max*(DF) parameter reaches the saturation region, the sensor can raise an alarm to indicate that pose recognition results may be inaccurate.

The new threshold value ensures the optimal balance between the minimization of the number of poses presented in true *dark* conditions classified as *bright* (False Negative) and true *bright* conditions classified as *dark* (False Positive). Poses presented in the conditions from the FN category are the most likely to be missed as the light pattern obtained by the PDs would be very flat (low ambient light level and weak shadow). Poses presented in the conditions from the FP category could be detected if the value of the baseline caused by the ambient light was lower than the intensity of the LEDs light reflected from a hand. Therefore, the utilization range of the active mode can depend on ambient light conditions and distance of a hand to the sensor. A more detailed investigation of this problem is an implication for future works.

The typical power utilization of the basic active gesture sensors is at the milliwatts order of magnitude like in the Okuli device, which can be reduced to circa 100 mW [8] or 3.78 mA in the partially open cavity package sensor [9].

Most of the basic passive sensors require usually roughly one to two orders of magnitude less current to operate. The 2x2 PIR sensor requires less than 50 μW for the operation [21]. The power consumption of the wireless signals utilizing AllSee sensor was measured for two types of prototypes. The ADC-based prototype uses 28.91 μW and analog-based one needs only 5.85 μW for the detection of 15 gestures per minute [18]. The average current consumption of a single PD of the optical linear gesture sensor is at the level of 1.1 mA. Taking into account that $D$=1.5%, the total current consumption of the PDs of the sensor is 132 μA (660 μW). In the active operating mode of the linear sensor, the total current was estimated at the level of 2.02 mA (10.1 mW), but the PD consumption was considered as the least favorable (1.7 mA) due to the catalogue note [27]. Having stated that the single PD current consumption is 1.1 mA, the active operating mode could utilize on average 1.982 mA (9.91 mW). Therefore, switching from the active to the passive operating mode leads to the reduction of the utilized power by 93.34%. The power utilization of the sensor can be reduced by application of different types of photodiodes as well as by designing the sensor for lower operating voltages.

Since switching between the active and the passive operating modes of the sensor relies only on the value of *max*(DF), the decision regarding the utilization of the LEDs does not require much computation. For the hardware utilized in the linear gesture sensor, the time required for the decision (switching condition) and additional sampling was measured. The time between the completions of the passive and active operating mode samplings (separated by the computation of the switching condition) was approximately 500μs. Therefore, the choice of the optimal operating mode while the gesture is performed should not affect the performance of the sensor as it operates with the 40Hz frequency (25ms time interval) [27].

## V. CONCLUSION

The accuracy of the hand pose recognition by the *pANN* classifier in favorable conditions was evaluated to be very high. Yet other orientations of the linear optical sensor were investigated as well. In the passive operating mode the sensor proved unable to operate properly not only in dark conditions (which is obvious), but also in strong light conditions (strong shadows, saturation of PDs). In this study, the maximum value of the DF was considered a good measure to decide when to switch between the passive and the active operating modes. Switching can be automatically performed whenever certain (learned) threshold value ($T_{max}$) is reached. It was also showed that to increase the accuracy of the pose detection, such threshold value could be shifted to delegate some uncertain lighting conditions to the active mode. This could slightly increase the power consumption but the accuracy could be also higher.

This work gives evidence that the passive operating mode of optical gesture sensors can be considered in a certain range of ambient light conditions instead of the more power hungry active operating mode. The very significant current demand reduction of the sensor in the passive mode can help mobile devices utilizing the gesture sensor, e.g. smart glasses, live for a longer time, without a significant reduction of the performance of the sensor.

In the future works, instead of the ensemble of classifiers, also one universal classifier could be trained for the data of both of the operating modes and compared with the remaining two, already trained. Yet another implication for further research would be to consider the high performance mode of the sensor relying on the sampling of both active and passive mode DF in one sampling cycle. Understanding the changing light conditions and possible accuracies of estimation of parameters such as longitudinal position, distance, pose category, etc., could be crucial for the automatic detection of different static and dynamic gestures.